\documentclass[12pt]{article}

\usepackage{epsfig}
\usepackage{amssymb,amsmath}
\usepackage{setspace}

\newcommand{\bea}{\begin{eqnarray}}
\newcommand{\eea}{\end{eqnarray}}

 \makeatletter
\def\alt{\mathrel{\mathpalette\gl@align<}}
\def\agt{\mathrel{\mathpalette\gl@align>}}
\def\gl@align#1#2{\lower.6ex\vbox{\baselineskip\z@skip\lineskip\z@
\ialign{$\m@th#1\hfil##\hfil$\crcr#2\crcr\sim\crcr}}} \makeatother

\begin{document}
%
\vspace*{1.0cm}

\begin{center}
\baselineskip 20pt {\Large\bf 
Higgs Inflation, Seesaw Physics and Fermion Dark Matter
}
\vspace{1cm}

{\large
Nobuchika Okada$^{a,}$\footnote{E-mail:okadan@ua.edu}
and Qaisar Shafi$^{b,}$\footnote{ E-mail:shafi@bartol.udel.edu}
} \vspace{.5cm}

{\baselineskip 20pt \it
$^b$Department of Physics and Astronomy,\\
University of Alabama, Tuscaloosa, AL 35487, USA \\
\vspace{2mm}
$^a$Bartol Research Institute, Department of Physics and Astronomy, \\
University of Delaware, Newark, DE 19716, USA 
}
\vspace{.5cm}

\vspace{1.5cm} {\bf Abstract}
\end{center}

We present an inflationary model in which the Standard Model Higgs doublet field with non-minimal coupling 
  to gravity drives inflation, and the effective Higgs potential is stabilized by new physics 
  which includes a dark matter particle and right-handed neutrinos for the seesaw mechanism. 
All of the new particles are fermions, so that the Higgs doublet is the unique inflaton candidate. 
With central values for the masses of the top quark and the Higgs boson, 
  the renormalization group improved Higgs potential is employed to yield 
  the scalar spectral index $n_s \simeq 0.968$, the tensor-to-scalar ratio $r \simeq 0.003$, and 
  the running of the spectral index $\alpha=dn_s/d \ln k \simeq -5.2 \times 10^{-4}$  for the number of e-folds $N_0=60$ 
  ($n_s \simeq 0.962$, $r \simeq 0.004$, and $\alpha \simeq -7.5 \times 10^{-4}$ for $N_0=50$).  
The fairly low value of $r \simeq 0.003$ predicted in this class of models means 
  that the ongoing space and land based experiments are not expected 
  to observe gravity waves generated during inflation.

\vspace{1.5cm} 
{\bf 
Dedicated to the memory of Dr.~Paul Weber (1947 - 2015). 
Paul was an exceptional human being and a very special friend who will be sorely missed.
}

\thispagestyle{empty}

\newpage

\addtocounter{page}{-1}

\baselineskip 18pt

\section{Introduction}

With the recent discovery of the Standard Model (SM) Higgs boson at the Large Hadron Collider (LHC), 
   it seems appropriate to reconsider whether the Higgs boson can successfully play the role of inflaton 
   in the early universe (Higgs inflation)~\cite{Higgs_inflation1, Higgs_inflation2, Higgs_inflation3}. 
Despite the presence of non-minimal coupling to gravity, which is a crucial ingredient, an important challenge in successfully 
   implementing SM Higgs inflaton has to do with the fact that the quartic Higgs coupling ($\lambda$) becomes negative 
   at an energy scale of order $10^{10}$ GeV~\cite{RGE_Higgs_quartic}. 
Without new physics, this can only be avoided by assuming values for the top quark pole mass that lie more than 4 sigmas 
   below the current world average of $173.34$ GeV~\cite{top_pole_mass}. 
Alternatively, an option not favored by experiments, for avoiding a negative quartic coupling is to assume values 
   for the Higgs boson mass that are somewhat larger than the current average $m_h \simeq125$ GeV.

In order to avoid the quartic Higgs coupling from turning negative at high energies, 
  we may simply introduce a real SM singlet scalar whose coupling with the Higgs doublet 
  can turn the beta function of the quartic Higgs coupling to be positive~\cite{Higgs_portal}.  
This scalar can be the so-called Higgs portal dark matter~\cite{Higgs_portal_DM}, 
  when a $Z_2$ parity is implemented to ensure the stability of the scalar.   
It has been recognized for sometime~\cite{instability_typeII, instability_typeIII} that the instability problem associated 
   with the quartic coupling can be overcome with new physics provided by type II~\cite{typeII} and type III~\cite{typeIII} 
   seesaw mechanisms which are often invoked  in understanding the solar and atmospheric neutrino oscillations.

In this paper, we consider Higgs inflation in the context of new physics 
  which not only solves the instability problem of the effective Higgs potential 
  but also supplements the SM with a dark matter candidate and the seesaw mechanism for neutrino masses. 
In our case, all new particles are fermions, so that  the SM Higgs field is the sole candidate for the role of the inflaton field.  
In the context of inflation there appears just a single new dimensionless parameter $\xi$, which is associated 
   with the non-minimal coupling of the Higgs scalar to gravity. 
With additional scalars, this would not be possible and the inflation could not be uniquely 
  identified with the SM Higgs field. 
Thus, a new scalar field can play the role of inflaton in the presence of a non-minimal gravitational coupling. 
For example, one may introduce a SM singlet scalar to drive inflation and yield the inflationary predictions 
  consistent with the observations~\cite{nm_phi4I, nm_phi4II}, with a lower bound $r > 0.002$ 
  for $ n_s \gtrsim 0.96$ when possible quantum corrections are taken into account~\cite{nm_phi4II}. 
This scalar may be identified as a B-L Higgs field in the minimal B-L model~\cite{nm_BL}.
Furthermore, the Higgs portal scalar dark matter can play the role of inflaton, 
  leading to a unification of inflaton and dark matter particle~\cite{inflatonDM}.  
For a scenario relating inflation, seesaw physics and Majoron dark matter, see Ref.~\cite{Majoron}.

The layout of this paper is as follows. 
In Sec.~2 we define our framework consisting of the SM supplemented by fermion dark matter and the seesaw mechanism, 
  and describe the renormalization group (RG) evolution of the quartic Higgs coupling.  
The parameter regions are identified to resolve the instability of the effective Higgs potential 
  and to reproduce the observed thermal dark matter relic abundance. 
With the RG improved effective potential and non-minimal gravitational coupling, 
   the Higgs inflation is analyzed and the inflationary predictions are presented in Sec.~3. 
As we previously mentioned, the magnitude of $r$ is estimated to be close to $0.003$ for $N_0=60$ 
  ($0.004$ for $N_0=50$),  which lies almost two orders of magnitude below the value reported 
  by the BICEP2 collaboration~\cite{BICEP2}. 
Our conclusions are summarized in Sec.~4.

\section{SM supplemented by new fermions}
For a fermion dark matter candidate, we consider a suitable multiplet of the SM SU(2) gauge group 
  with an appropriate hypercharge to include an electrically neutral fermion 
  as a candidate for dark matter (minimal dark matter)~\cite{MDM}.   
A $Z_2$ parity is introduced, and an odd parity is assigned to the multiplet so as to ensure 
  the stability of the dark matter candidate. 
A variety of SU(2) multiplets have been considered in \cite{MDM}, 
  where the properties of the dark matter candidate are summarized in Table~1~\cite{MDM}. 
Because of the SM gauge invariance and the $Z_2$ parity, 
  the fermion dark matter only has electroweak interaction, 
  and the dark matter properties are completely determined by its mass. 
Through the electroweak interaction, the observed thermal relic abundance 
  is reproduced with the dark matter mass around a TeV.  
Among a variety of choices for the dark matter candidate, we consider in this paper two simple cases:
   (1) an SU(2) triplet with zero hypercharge, and  (2) a {\bf 5}-plet of SU(2) with zero hypercharge. 
Note that we have chosen the multiplets with zero hypercharge to avoid severe constraints 
   from the direct dark matter search experiments~\cite{MDM}. 
Although these multiplets have no direct coupling with the Higgs doublet,  
   they effectively contribute to the beta function of the quartic Higgs coupling 
   through the running SU(2) gauge coupling. 
In the presence of the SU(2) multiplets, the running SU(2) gauge coupling is altered 
   to be asymptotically non-free and to yield larger positive contributions to the beta function 
   of the quartic Higgs coupling than those in the SM.  
As a result, the instability problem can be solved if the positive contribution is large enough 
   or at least, the problem becomes milder.

In order to naturally incorporate non-zero neutrino masses, we consider type I~\cite{typeI} 
  or type III~\cite{typeIII} seesaw, where SM singlet or SU(2) triplet right-handed neutrinos, respectively, are introduced.  
Since, like the top quark Yukawa coupling, the Dirac neutrino Yukawa couplings 
  yield a negative contribution to the beta function of the quartic Higgs coupling, 
  a large Dirac Yukawa coupling makes the situation worse for the instability problem.  
If the Dirac Yukawa coupling is negligibly small, the right-handed neutrinos in type I seesaw 
  have no effect on the RGE analysis (at the 1-loop level).  
The SU(2) triplet neutrinos in type III seesaw work to prevent the running quartic Higgs coupling from becoming negative.   
It has been shown in \cite{ instability_typeIII} that type III seesaw with TeV or lower seesaw scale 
  can solve the instability problem.  
Such light SU(2) triplet neutrinos can be tested at the LHC Run II with a collider energy of 13--14 TeV.

Let us now define our representative models consisting of the SM supplemented 
   by the dark matter candidate and type I/III seesaw. 
We may combine cases (1) and (2) for the dark matter candidate with type I and/or type III seesaw. 
As we will see in the following analysis, the triplet dark matter in case (1) needs to be combined 
  with type III seesaw to solve the instability problem. 
On the other hand, once the {\bf 5}-plet dark matter is introduced, 
  the effective Higgs potential becomes stable. 
Hence, as simple examples, we consider the following two cases: 
(i) SM supplemented by SU(2) triplet dark matter and type III seesaw,  
and  
(ii) SM supplemented by an SU(2) {\bf 5}-plet dark matter and type I seesaw.

\subsection{Case (i)}
In order to reproduce the observed thermal relic abundance, 
  the mass of the triplet dark matter is set to be $M_{DM}=2.4$ TeV~\cite{MDM}. 
Since two generations of right-handed neutrinos are sufficient to reproduce the neutrino oscillation data, 
  we introduce, for simplicity, two SU(2) triplet neutrinos with a degenerate mass $M_R$.  
For renormalization scale $\mu < M_R$ and $M_{DM}$,  
  the dark matter and right-handed neutrinos are decoupled, 
  and we employ the SM RG equations at two-loop level~\cite{RGE}.  
For the three SM gauge couplings $g_i$ ($i=1,2,3$), we have 
\bea
 \frac{d g_i}{d \ln \mu} =
 \frac{b_i}{16 \pi^2} g_i^3 +\frac{g_i^3}{(16\pi^2)^2}
  \left( \sum_{j=1}^3 B_{ij}g_j^2 - C_i y_t^2   \right),
\eea
 where
\bea
b_i = \left(\frac{41}{10},-\frac{19}{6},-7\right),~~~~
 { B_{ij}} =
 \left(
  \begin{array}{ccc}
  \frac{199}{50}& \frac{27}{10}&\frac{44}{5}\\
 \frac{9}{10} & \frac{35}{6}&12 \\
 \frac{11}{10}&\frac{9}{2}&-26
  \end{array}
 \right),  ~~~~
C_i=\left( \frac{17}{10}, \frac{3}{2}, 2 \right),
\eea 
and among the SM Yukawa couplings, only the top Yukawa coupling ($y_t$) is included in our analysis.  
The RG equation for the top Yukawa coupling given by
\bea 
 \frac{d y_t}{d \ln \mu}
 = y_t  \left(
 \frac{1}{16 \pi^2} \beta_t^{(1)} + \frac{1}{(16 \pi^2)^2} \beta_t^{(2)}
 \right), 
\eea
where the one-loop contribution is
\bea
 \beta_t^{(1)} =  \frac{9}{2} y_t^2 -
  \left(
    \frac{17}{20} g_1^2 + \frac{9}{4} g_2^2 + 8 g_3^2
  \right) ,
\eea
while the two-loop contribution is given by 
\bea
\beta_t^{(2)} &=&
 -12 y_t^4 +   \left(
    \frac{393}{80} g_1^2 + \frac{225}{16} g_2^2  + 36 g_3^2
   \right)  y_t^2  \nonumber \\
 &&+ \frac{1187}{600} g_1^4 - \frac{9}{20} g_1^2 g_2^2 +
  \frac{19}{15} g_1^2 g_3^2
  - \frac{23}{4}  g_2^4  + 9  g_2^2 g_3^2  - 108 g_3^4 \nonumber \\
 &&+ \frac{3}{2} \lambda^2 - 6 \lambda y_t^2 .
\eea
The RG equation for the quartic Higgs coupling is given by 
\bea
\frac{d \lambda}{d \ln \mu}
 =   \frac{1}{16 \pi^2} \beta_\lambda^{(1)}
   + \frac{1}{(16 \pi^2)^2}  \beta_\lambda^{(2)},
\eea
with
\bea
 \beta_\lambda^{(1)} &=& 12 \lambda^2 -
 \left(  \frac{9}{5} g_1^2+ 9 g_2^2  \right) \lambda
 + \frac{9}{4}  \left(
 \frac{3}{25} g_1^4 + \frac{2}{5} g_1^2 g_2^2 +g_2^4
 \right) + 12 y_t^2 \lambda  - 12 y_t^4 ,
\eea
and
\bea
  \beta_\lambda^{(2)} &=&
 -78 \lambda^3  + 18 \left( \frac{3}{5} g_1^2 + 3 g_2^2 \right) \lambda^2
 - \left( \frac{73}{8} g_2^4  - \frac{117}{20} g_1^2 g_2^2
 - \frac{1887}{200} g_1^4  \right) \lambda - 3 \lambda y_t^4
 \nonumber \\
 &&+ \frac{305}{8} g_2^6 - \frac{289}{40} g_1^2 g_2^4
 - \frac{1677}{200} g_1^4 g_2^2 - \frac{3411}{1000} g_1^6
 - 64 g_3^2 y_t^4 - \frac{16}{5} g_1^2 y_t^4
 - \frac{9}{2} g_2^4 y_t^2
 \nonumber \\
 && + 10 \lambda \left(
  \frac{17}{20} g_1^2 + \frac{9}{4} g_2^2 + 8 g_3^2 \right) y_t^2
 -\frac{3}{5} g_1^2 \left(\frac{57}{10} g_1^2 - 21 g_2^2 \right)
  y_t^2  - 72 \lambda^2 y_t^2  + 60 y_t^6.
\eea
In solving the RGEs, we use the boundary conditions at the top quark pole mass ($M_t$)
  given in \cite{RGE_Higgs_quartic}: 
\bea 
g_1(M_t)&=&\sqrt{\frac{5}{3}} \left( 0.35761 + 0.00011 (M_t - 173.10) 
   -0.00021  \left( \frac{M_W-80.384}{0.014} \right) \right),  
\nonumber \\
g_2(M_t)&=& 0.64822 + 0.00004 (M_t - 173.10) + 0.00011  \left( \frac{M_W-80.384}{0.014} \right) , 
\nonumber \\
g_3(M_t)&=& 1.1666 + 0.00314 \left(  \frac{\alpha_s-0.1184}{0.0007}   \right) ,
\nonumber \\
y_t(M_t) &=& 0.93558 + 0.0055 (Mt - 173.10) - 0.00042  \left(  \frac{\alpha_s-0.1184}{0.0007}   \right)
  - 0.00042 \left( \frac{M_W-80.384}{0.014} \right) ,
\nonumber \\
\lambda(M_t) &=&  2 (0.12711 + 0.00206 (m_h - 125.66) - 0.00004 (M_t - 173.10) ) .
\eea 
We employ $M_W=80.384$ (in GeV), $\alpha_s =0.1184$, 
   the central value of the combination of Tevatron and LHC measurements of top quark mass 
   $M_t=173.34$ (in GeV)~\cite{top_pole_mass}, and the central value of the updated Higgs boson mass measurement, 
   $m_h=125.03$ (in GeV) from the CMS experiment~\cite{Higgs_mass_CMS}, for example.\footnote{
Instead of  the CMS result, one may use the result of the ATLAS experiment~\cite{Higgs_mass_ATLAS}. 
The difference between our results using the CMS and ATLAS data is negligibly small. 
}

For the renormalization scale $\mu \geq M_R$ and/or $M_{DM}$,  
   the SM RG equations should be modified to include contributions from the new particles 
   and, in particular, the RG evolution of the quartic Higgs coupling is altered. 
In this paper, we take only one-loop corrections from the new particles into account. 
As we will see, the quartic Higgs coupling is prevented from becoming negative for $M_R \lesssim 6$ TeV,  
   and in this case the Dirac Yukawa coupling in type III seesaw is negligibly small. 
Hence, the new particles effectively modify only the beta function of the SU(2) gauge coupling. 
For $\mu \geq M_R$, the beta function coefficient of the SU(2) gauge coupling receives 
  a new contribution from the 2-generations of right-handed neutrinos given by $\Delta b_2 = \frac{4}{3} \times 2$, 
  while the new contribution from the triplet dark matter multiplet is given by $\Delta b_2({\rm DM}) = \frac{4}{3}$ 
  for $\mu  \geq M_{DM}=2.4$ TeV. 
 
\begin{figure}[ht]
  \begin{center}
   \includegraphics[width=10cm]{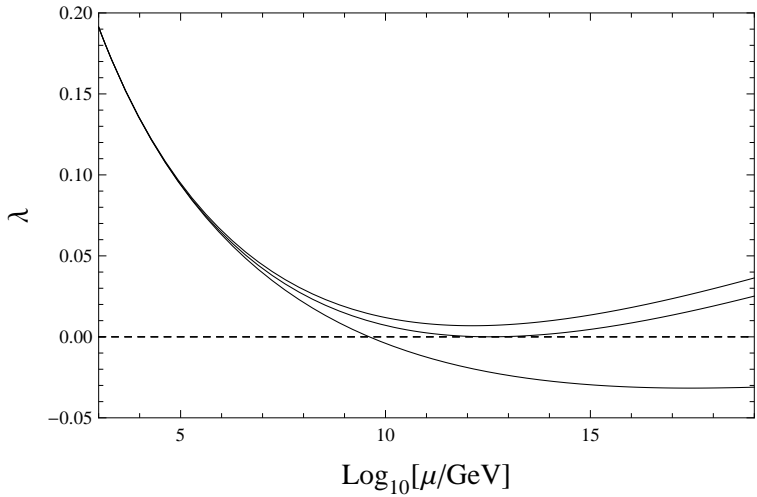}
   \end{center}
\caption{
RG evolution of the quartic Higgs coupling for $M_R=250$ GeV and $5750$ GeV  
  along with the SM case (solid lines from top to bottom). 
We set $M_{DM}=2.4$ TeV  to reproduce the observed dark matter relic density. 
}
  \label{fig:RGE-SS3}
\end{figure}

In Fig.~\ref{fig:RGE-SS3}, we show the running of the quartic Higgs coupling for $M_R=250$ GeV and $5750$ GeV  
  along with the SM result as solid lines from top to bottom. 
Here we set $M_{DM}=2.4$ TeV  to reproduce the observed dark matter relic density. 
The quartic Higgs coupling is kept positive below the reduced Planck mass $M_P=2.435 \times 10^{18}$ GeV 
  for $M_R < 5750$ GeV.  
There are lower bounds from the search for type III seesaw right-handed neutrino at the LHC.  
The ATLAS experiment~\cite{ATLAS_type3} has set the lower bound $M_R > 245$ GeV at 95\% CL, 
  and the CMS experiment~\cite{CMS_type3} has given a similar bound, $M_R > 180-210$ GeV.

\subsection{Case (ii)}
In this case, we introduce fermion dark matter belonging to a {\bf 5}-plet of SU(2) with zero hypercharge. 
We set the dark matter mass to be $M_{DM}=4.4$  TeV to reproduce the observed relic abundance~\cite{MDM}. 
For the renormalization scale $\mu > M_{DM}$, the beta function coefficient receives a new contribution 
  from the dark matter multiplet, $\Delta b_2 = \frac{20}{3}$, which is large enough to prevent the running 
  quartic Higgs coupling from turning negative for $\mu < M_P$.

For the neutrino mass generation in type I seesaw, we introduce SM gauge singlet right-handed neutrinos 
   $\psi_i$, where $i$ is a generation index.  
The relevant terms in the Lagrangian are given by 
\bea 
 {\cal L} \supset 
 - y_{ij} \overline{\ell_i} \psi_j H - M_R^{ij} \overline{\psi^c_i} \psi_j ,   
\eea
 where $\ell_i$ is the $i$-th generation SM lepton doublet, and $M_R$ 
  is a Majorana mass matrix for the right-handed neutrinos.  
Integrating out  the right-handed neutrinos at energies below $M_R$, 
  the effective dimension five operator is generated by the seesaw mechanism. 
After the electroweak symmetry breaking, the light neutrino mass matrix is obtained as 
\bea
  {\bf M}_\nu = 
  \frac{v^2}{2} {\bf Y}_\nu^T M_R^{-1} {\bf Y}_\nu ,
\eea
  where $v=246$ GeV is the vacuum expectation value of the Higgs doublet,
  and ${\bf Y}_\nu = y_{ij}$ is a 3$\times$3 Yukawa matrix. 
Since type I seesaw involves many free parameters, 
  we assume that the heaviest right-handed neutrino provides a dominant impact 
  on the RG evolution of the quartic Higgs coupling. 
Hence, the relevant term is simplified with one right-handed neutrino as 
\bea 
 {\cal L} \supset 
 - y_D \overline{\ell} \psi H - M_R \overline{\psi^c} \psi ,   
\eea
with a Dirac Yukawa coupling $y_D$ and a right-handed neutrino mass $M_R$.

For the renormalization scale $\mu \geq M_R$, the SM RG equations are modified 
  in the presence of the right-handed neutrino as 
\bea 
&& \beta_t^{(1)} \to \beta_t^{(1)}  + S_\nu, 
\nonumber \\
&& \beta_{\lambda}^{(1)} \to \beta_{\lambda}^{(1)} 
  + 4  S_\nu \lambda - 4 S_\nu^2,
\eea
where $S_\nu=y_D^\dagger y_D$, and its corresponding RG equation is given by 
\bea
16 \pi^2 \frac{d S_\nu}{d \ln \mu} 
 = S_\nu  \left[ 6 y_t^2 + 5 S_\nu  -\left( \frac{9}{10} g_1^2 +\frac{9}{2} g_2^2 \right) \right] .
\eea
To yield the mass scale of the neutrino oscillation data for the atmospheric neutrino 
  through the seesaw mechanism, we fix the relation between $S_\nu$ and $M_R$ 
  by $S_\nu v^2/(2 M_R) = 10^{-10}$ GeV. 
Thus, the Dirac Yukawa coupling can be large,  $S_\nu ={\cal O}(1)$, for $M_R={\cal O}(10^{14}\;{\rm GeV})$. 
In this case, the effect of type I seesaw on the RG evolution of the quartic Higgs coupling  
  can be significant enough to make the effective Higgs potential unstable.

\begin{figure}[ht]
  \begin{center}
   \includegraphics[width=10cm]{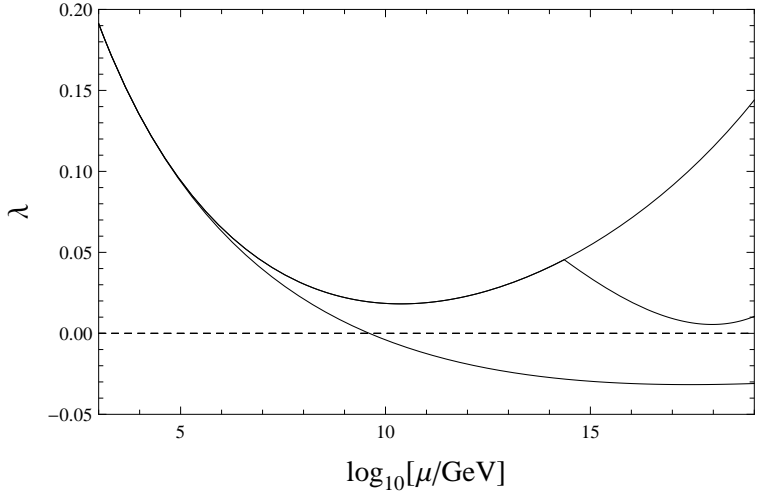}
  \end{center}
  \caption{
RG evolution of the quartic Higgs coupling
  for $M_R \ll 10^{14}$ GeV and $M_R =2.3 \times 10^{14}$ GeV, 
  along with the SM case (solid lines from top to bottom). 
We have fixed $M_{DM}=4.4$ TeV  to reproduce the observed dark matter relic density. 
}
  \label{fig:RGE-SS1}
\end{figure}

In Fig.~\ref{fig:RGE-SS1}, we show the RG evolutions of the quartic Higgs coupling 
  for $M_R \ll 10^{14}$ GeV and $M_R =2.3 \times 10^{14}$ GeV, 
  along with the SM case (solid lines from top to bottom). 
Here we have fixed $M_{DM}=4.4$ TeV. 
The top solid line shows that the effective Higgs potential becomes stable 
  in the presence of the SU(2) {\bf 5}-plet.  
When type I seesaw effect is significant (middle solid line), the beta function to the quartic Higgs coupling
  becomes negative at $\mu=M_R$,  and the RG evolution shows a minimum 
  ($\lambda_{\rm min} \simeq 5.5 \times 10^{-3}$)  at $\mu \simeq M_P$.

\section{Running Higgs inflation}
In the SM supplemented by the dark matter particle and type I or III seesaw, 
   the quartic Higgs coupling stays positive below the Planck scale 
   and the instability of the effective Higgs potential is resolved. 
Employing the effective Higgs potential, we consider Higgs inflation by introducing 
  a non-minimal gravitational coupling between the SM Higgs doublet and the scalar curvature. 
The basic action in the Jordan frame is given by 
\bea 
S_J =   \int d^4 x \sqrt{-g} 
 \left[
- \left( \frac{1}{2} + \xi H^\dagger H \right) \mathcal{R}
+( {\cal D}_\mu H)^\dagger ( {\cal D}_\mu H)
- \frac{1}{2} \lambda \left(  H^\dagger H - \frac{v^2}{2} \right)^2  
\right],
\eea
where we set the Planck scale $M_P=1$, and 
  $H = (0, v + \phi )/\sqrt{2}$ is the Higgs doublet field in the unitary gauge 
  with the physical Higgs ($\phi$) identified as the inflaton.  
In the following analysis, we employ the RG improved effective inflaton potential given by 
\bea 
 V(\phi) = \frac{\lambda(\phi)}{8} \phi^4, 
\eea
where $\lambda(\phi)$ is the solution to the RG equation for the quartic Higgs coupling 
  with the identification $\mu=\phi$, and we only consider the quartic coupling during inflation.

In the Einstein frame with a canonical gravity sector, 
  we describe the theory with a new inflaton field ($\sigma$) which has a canonical kinetic term. 
The relation between $\sigma $ and $\phi$ is given by
\bea
\left({d\sigma \over d\phi}\right)^{-2} =
 \frac{\left( 1 + \xi \phi^2 \right)^2}{1+(6 \xi +1) \xi \phi^2} \; .
\eea
The action in the Einstein frame is then given by
\bea
S_E = \int d^4 x \sqrt{-g_E}\left[-\frac12  \mathcal{R}_E+\frac12 (\partial \sigma)^2
-V_E(\phi(\sigma)) \right],
\eea
with~\cite{VE}
\bea
V_E(\phi) = \frac{\lambda(\Phi)}{8} \Phi^4,  
\eea
where $ \Phi = \phi/\sqrt{1+\xi \phi^2} $.

The inflationary slow-roll parameters in terms of the original 
 scalar field ($\phi$) are expressed as 
\bea
\epsilon(\phi)&=&\frac{1}{2} \left(\frac{V_E'}{V_E \sigma'}\right)^2, 
 \nonumber \\
\eta(\phi)&=& 
\frac{V_E''}{V_E (\sigma')^2}- \frac{V_E'\sigma''}{V_E (\sigma')^3} ,  
 \nonumber \\
\zeta (\phi) &=&  \left(\frac{V_E'}{V_E \sigma'}\right) 
 \left( \frac{V_E'''}{V_E (\sigma')^3}
-3 \frac{V_E'' \sigma''}{V_E (\sigma')^4} 
+ 3 \frac{V_E' (\sigma'')^2}{V_E (\sigma')^5} 
- \frac{V_E' \sigma'''}{V_E (\sigma')^4} \right)  , 
\eea
where a prime denotes a derivative with respect to $\phi$. 
The amplitude of the curvature perturbation $\Delta_{\mathcal{R}}$ is given by 
\begin{equation} 
\Delta_{\mathcal{R}}^2 = \left. \frac{V_E}{24 \pi^2 \epsilon } \right|_{k_0},
\end{equation}
  which should satisfy $\Delta_\mathcal{R}^2= 2.215\times10^{-9}$
  from the Planck measurement \cite{Planck2013} 
  with the pivot scale chosen at $k_0 = 0.05$ Mpc$^{-1}$.
  The number of e-folds is given by
\begin{eqnarray}
  N_0 = \frac{1}{2} \int_{\phi_{\rm e}}^{\phi_0}
\frac{d\phi}{\sqrt{\epsilon(\phi)}}\left(\frac{d\sigma}{d\phi}\right)  ,
\end{eqnarray} 
where $\phi_0$ is the inflaton value at horizon exit of the scale corresponding to $k_0$, 
  and $\phi_e$ is the inflaton value at the end of inflation, 
  which is defined by ${\rm max}[\epsilon(\phi_e), | \eta(\phi_e)| ]=1$.
The value of $N_0$ depends logarithmically on the energy scale during inflation 
  as well as on the reheating temperature, and is typically around 50--60.

The slow-roll approximation is valid as long as the conditions 
   $\epsilon \ll 1$, $|\eta| \ll 1$ and $\zeta \ll 1$ hold. 
In this case, the inflationary predictions, 
   the scalar spectral index $n_{s}$, the tensor-to-scalar ratio $r$, 
   and the running of the spectral index $\alpha=\frac{d n_{s}}{d \ln k}$, are given by
\bea
n_s = 1-6\epsilon+2\eta, \; \; 
r = 16 \epsilon,  \; \;
\alpha=16 \epsilon \eta - 24 \epsilon^2 - 2 \zeta. 
\eea 
Here the slow-roll parameters are evaluated at $\phi=\phi_0$.

\begin{table}[ht]
\begin{center}
\begin{tabular}{|c|ccccc|}
\hline
& \multicolumn{5}{|c|}{Case (i)}   \\
\hline 
 $N_0$ & $M_R ({\rm GeV})$ &  $ \xi $    &  $n_s$       &  $r$               &  $-\alpha (10^{-4})$  \\ 
\hline
      50  &   250                         &   4320    &  $ 0.962 $  & $0.00421$    &  $7.50  $              \\
            & 5750                         &   3192    &  $ 0.962 $  & $0.00421  $  &  $7.50  $             \\
\hline  
      60  &   250                         &   5124     &  $ 0.968 $  & $0.00297$    &  $5.24  $              \\    
            & 5750                         &   3776     &  $ 0.968 $  & $0.00297$    &  $5.24  $              \\     
\hline\hline
        &  \multicolumn{5}{|c|}{Case (ii)}  \\ 
\hline
 $N_0$ & $M_R (10^{14}\; {\rm GeV})$ & $ \xi $ &  $n_s$       &  $r$             &  $-\alpha (10^{-4})$   \\
\hline \hline
50         &  $ \ll 1$                                    &  8361  &  $ 0.962 $  & $0.00421$  &  $7.50  $    \\
             &  $2.3  $                                    &  3311  &  $ 0.962 $  & $0.00416 $   &  $7.45 $   \\    
\hline 
60         &  $ \ll 1$                                    &  9921  &  $  0.968 $  & $0.00297  $   &  $5.24  $   \\   
             &  $2.3  $                                    & 4008   &  $ 0.968 $  & $0.00294  $   &  $5.21  $ \\  
\hline
\end{tabular}
\end{center}
\caption{ 
Inflationary predictions in Case (i) and Case (ii) for $N_0=50$ and $60$. 
} 
\end{table}

For Case (i) and Case (ii) discussed in the previous section, 
  we calculate the inflationary prediction with the RG improved effective potential. 
The results for the cases presented in Figs.~1 and 2 are summarized in Table 1. 
For a fixed $N_0$ value, the inflationary predictions are almost the same, 
  while the non-minimal coupling varies to satisfy $\Delta_\mathcal{R}^2= 2.215\times10^{-9}$
  from the Planck measurement.

\begin{figure}[ht]
  \begin{center}
   \includegraphics[width=10cm]{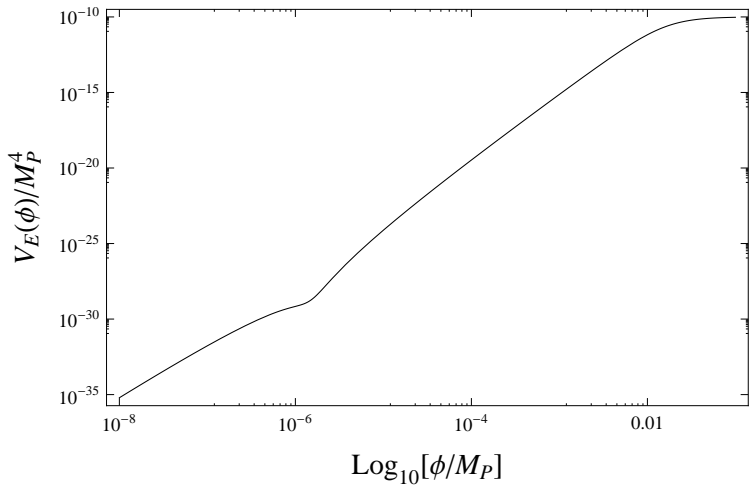}
   \end{center}
  \caption{
The RGE improved effective Higgs potential in the Einstein frame for $M_R=5750$ GeV 
  and $\xi =3776$ in Case (i). 
The effective potential shows an inflection point at $\phi/M_P \simeq 10^{-6}$, 
   which corresponds to the minimal value of the running quartic coupling, 
   $\lambda_{\rm min} \simeq 2.8 \times 10^{-5}$ in Fig.~\ref{fig:RGE-SS3}. 
}
  \label{fig:Veff}
\end{figure}

In Fig.~\ref{fig:RGE-SS3}, the RG evolution for $M_R=5750$ GeV (middle solid line) shows a minimum 
  ($\lambda_{\rm min} \simeq 2.8 \times 10^{-5}$) at $\mu \simeq 10^{12}$ GeV.  
For this case with $\xi =3776$ (see Table 1), 
  the RG improved effective Higgs potential in the Einstein frame is depicted in Fig.~\ref{fig:Veff}. 
We can see that the effective potential shows an inflection point at $\phi/M_P \simeq 10^{-6}$, 
   corresponding to  $\lambda_{\rm min}$. 
As $M_R$ is slightly raised (while keeping $\lambda_{\min} > 0$), 
   a local minimum in the effective potential develops  at $\phi/M_P \simeq 10^{-6}$, 
   so that the inflaton field will be trapped in this minimum after inflation. 
A second inflation then takes place until the vacuum transition from this local minimum 
   to the true electroweak vacuum. 
The condition to avoid this problem is stronger than the condition for the vacuum stability, 
  $\lambda (\mu) > 0$ for $\mu > M_P$. 
  From the RG evolution shown in Fig.~\ref{fig:RGE-SS1} (middle solid line), 
  we expect that the same problem occurs in Case (ii) 
  when the Dirac Yukawa is raised to make the minimum value of the running $\lambda$ very close to zero. 
However, the analysis of the effective potential in this case is more complicated, 
  since the scale corresponding to $\lambda_{\rm min}$ is close to the initial inflaton value, 
  and hence the change of the effective potential shape from varying $M_R$ 
  directly affects the inflationary predictions.

It has been shown in Ref.~\cite{critical} that if we fine-tune the input top quark mass $M_t \simeq 171$ GeV 
  to realize $\lambda_{\min} \simeq 10^{-6}$  at $\mu \simeq M_P$,  
  the effective Higgs potential develops an inflection point like the one shown in Fig.~\ref{fig:Veff}, 
  but at the Planck scale, and the Higgs inflation can predict $n_s \simeq 0.96$ and $ r \simeq 0.1$. 
This prediction for the tensor-to-scalar ratio is compatible with the BICEP2 result, 
  while Higgs inflation normally predicts $r \ll 0.1$ as we have shown in Table~1. 
Similar results have been obtained in an extension of the SM with a new scalar field  
  to avoid the instability problem~\cite{critical_BSM}, when model parameters are fine-tuned  
  to realize an inflection point in the effective Higgs potential at $\mu \simeq M_P$.  
Since we can realize a similar situation in Case (ii) by tuning $M_R$ values, 
  our Higgs inflation scenario might be able to predict $r \simeq 0.1$ for finely tuned input parameters.  
However, in order to show this, a very delicate analysis with fine-tunings 
  of multiple free parameters ($y_D$, $M_R$ and $\xi$) is necessary, and we leave it for future work.

\section{Conclusions}
The long-sought-after Higgs boson has been discovered at the LHC, and 
   this marks the beginning of the experimental confirmation of the SM Higgs sector.  
The observed Higgs boson mass of $\simeq 125$ GeV indicates that the electroweak vacuum is unstable, 
  since the quartic Higgs coupling becomes negative far below the Planck mass, 
  assuming that $M_t \simeq 173$ GeV.   
This instability problem has a great impact on the Higgs inflation scenario, 
  since the effective Higgs potential is no longer suitable for inflation. 
In order to solve the instability problem, we need some new physics which can alter 
  the RG evolution of the quartic Higgs coupling and keep the running coupling positive during inflation.

To realize Higgs inflation with the 125 GeV mass, we have supplemented the SM 
   with dark matter candidates and type I/III seesaw.  
A crucial point in introducing new particles is to retain the original idea of Higgs inflation, 
  namely, the SM Higgs field is the unique candidate for inflaton. 
Therefore, all new particles must be fermions. 
With this requirement, we have introduced fermion dark matter and right-handed neutrinos 
  for the seesaw mechanism.  
Two major missing pieces in the SM, a dark matter particle and neutrino masses, have been resolved in our scenario. 
The SM Higgs field with non-minimal gravitational coupling drives inflation 
  as the unique candidate for inflaton. 
Employing the effective Higgs potential with new particle contributions, 
  we have found the inflationary prediction of the scenario as 
   $n_s=0.968$, $r=0.003$ and $\alpha=-0.00052$ for $N_0=60$ 
  ($n_s=0.962$, $r=0.004$ and $\alpha=-0.00075$ for $N_0=50$), 
  which are consistent with the Planck measurements. 
With $r \ll 0.1$, this scenario will be excluded if the recently reported BICEP2 results are verified by ongoing experiments.

\section*{Acknowledgments}
Q.S. acknowledges support provided by the DOE grant No. DE-FG02-12ER41808. 


\end{document}